\documentclass[11pt]{article}
\usepackage{cite}
\usepackage[a4paper,left=0.5in,right=1in]{geometry}
\usepackage[usenames]{color}
\usepackage{graphicx}
\usepackage{epsfig}
\usepackage{dcolumn}
\usepackage{bm}

\begin{document}

\def \d {{\rm d}}
\def \q {Q}
\def \etta {F} 
\def \t {{\Theta}}
\def \k {{\kappa}}
\def \l {{\lambda}}
\def \s {{\sigma}}
\def \tr {{\tilde\rho}}
\def \tv {{\tilde v}}
\def \tz {{\tilde z}}
\def \e {{\epsilon}}
\def \P {{p_\lambda}}
\def \Q {{p_\nu}}
\def \eigen {{\mathcal N}}
\def \M {{\mathcal M}}
\def \bm #1 {\mbox{\boldmath{$m^{(#1)}$}}}

\newcommand{\be}{\begin{equation}}
\newcommand{\ee}{\end{equation}}

\newcommand{\beqn}{\begin{eqnarray}}
\newcommand{\eeqn}{\end{eqnarray}}
\newcommand{\AdS}{anti--de~Sitter }
\newcommand{\AAdS}{\mbox{(anti--)}de~Sitter }
\newcommand{\AAN}{\mbox{(anti--)}Nariai }
\newcommand{\AS}{Aichelburg-Sexl }
\newcommand{\pa}{\partial}
\newcommand{\pp}{{\it pp\,}-}
\newcommand{\ba}{\begin{array}}
\newcommand{\ea}{\end{array}}

\title{Robinson--Trautman spacetimes with an electromagnetic field in higher dimensions}

\author{Marcello Ortaggio\thanks{ortaggio`AT'ffn.ub.es} \\
 Departament de F\'{\i}sica Fonamental, \\ 
 Universitat de Barcelona, Diagonal 647, E-08028 Barcelona, Spain 
 \\ \\
 Ji\v{r}\'{\i} Podolsk\'y\thanks{podolsky`AT'mbox.troja.mff.cuni.cz} \ and Martin \v{Z}ofka\thanks{zofka`AT'mbox.troja.mff.cuni.cz}  
\\ Institute of Theoretical Physics, Faculty of Mathematics and Physics,\\
 Charles University in Prague, V Hole\v{s}ovi\v{c}k\'{a}ch 2, 180 00 Prague 8,  Czech Republic}

\date{\today}

\maketitle

\abstract{We investigate higher dimensional Robinson--Trautman spacetimes with an electromagnetic field aligned with the hypersurface orthogonal, non-shearing, expanding geodesic null congruence. After integrating the system of Einstein--Maxwell equations with an arbitrary cosmological constant, we present the complete family of solutions. In  {\em odd} spacetime dimensions they represent (generalized) Reissner--Nordstr\" om--de~Sitter black holes. The event horizon (more generically, the transverse space) may be any Einstein space, and the full metric is specified by three independent parameters related to mass, electric charge and cosmological constant. These solutions also exhaust the class of Robinson--Trautman spacetimes with an aligned Maxwell--Chern--Simons field (the CS term must vanish because of the alignment assumption and of the Einstein equations). In {\em even} dimensions an additional magnetic ``monopole-like'' parameter is also allowed provided now the transverse space is an (almost-)K\"ahler Einstein manifold. The Weyl tensor of all such solutions is of algebraic type D.  We also consider the possible inclusion of aligned pure radiation.

}

\vspace{.2cm}





\section{Introduction}
\label{intro}

In General Relativity, the study of ray optics has played a major
role in the construction, intepretation and invariant classification of exact solutions  (see, e.g., \cite{Stephanibook} for a review and for original references). This applies in particular to solutions representing gravitational radiation. During the Golden Age of theoretical studies of exact radiative spacetimes, Robinson and Trautman introduced and investigated ${D=4}$ dimensional Lorentzian geometries that admit a geodesic, non-twisting, non-shearing, expanding null congruence \cite{RobTra60,RobTra62}. The Robinson--Trautman family is by now one of the fundamental classes of exact solutions to Einstein's field equations in vacuum and with principal matter fields such as pure radiation or an electromagnetic field \cite{Stephanibook}. It includes a number of well-known spacetimes ranging from static black holes and the Vaidya solution to the C-metric and other radiative solutions. Noticeably, the Goldberg--Sachs theorem \cite{Stephanibook} implies that Robinson--Trautman  geometries are algebraically special (at least in vacuum and with ``sufficiently aligned'' matter fields), since they are non-shearing. In fact, explicit vacuum solutions of all 
special Petrov types are known \cite{RobTra60,RobTra62,Stephanibook}.

The geometric optics approach was naturally developed in the framework of ${D=4}$ General Relativity. On the other hand, in recent years string theory and specific extra-dimension scenarios have stimulated the investigation of gravity in more than four spacetime dimensions. It is thus now interesting to
consider possible extensions of the above concepts to arbitrary (higher) dimensions, and their relation to the ${D>4}$ classification of the Weyl tensor \cite{Coleyetal04}. In \cite{FroSto03,Pravdaetal04,LewPaw05,OrtPraPra07,PraPraOrt07}, various general aspects of geometric optics in ${D>4}$ dimensions (which evades the standard $D=4$ Goldberg--Sachs theorem in many ways) have been analyzed. In \cite{PodOrt06}, the Robinson--Trautman family of solutions has been extended to higher dimensions in the case of empty space possibly with a  cosmological constant and in the case of aligned pure radiation. The authors pointed out important differences with respect to the ${D=4}$ case for vacuum spacetimes (see also \cite{Ortaggio07,PraPraOrt07}). However, from a higher dimensional perspective one would also be interested in theories that incorporate electromagnetic fields. It is thus the purpose of this paper to study Robinson--Trautman spacetimes in the higher dimensional Einstein--Maxwell theory (for any value of the cosmological constant). For simplicity, we will focus on aligned fields. In ${D\ge 5}$ odd dimensions, we shall also consider the inclusion of an additional Chern--Simons term, which gives rise, e.g., to the bosonic sector of five-dimensional
minimal (gauged) supergravity.

The paper is organized as follows. In section~\ref{sec_geom} we present the line element of generic Robinson--Trautman spacetimes \cite{PodOrt06} and we study purely algebraic properties of an aligned Maxwell field. In section~\ref{Einstein} we proceed by integrating systematically the full set of Einstein--Maxwell equations within such a setting. We summarize the obtained spacetimes and we discuss some special cases in section~\ref{sec_discussion}. Concluding remarks are in given section~\ref{sec_conclusions}. Throughout the paper we focus on $D>4$ dimensions, and well-known results in the special case $D=4$ are summarized in the Appendix.

\section{Robinson--Trautman geometry and aligned Maxwell fields}
\label{sec_geom}

As shown in \cite{PodOrt06}, the general line element for any $D$-dimensional spacetime which admits a non-twisting, non-shearing but expanding congruence \cite{FroSto03,Pravdaetal04} generated by the geodesic null vector field~{\boldmath $k$} can be written as
\begin{equation}\label{geo_metric}
  \d s^2=g_{ij}\left(\d x^i+ g^{ri}\d u\right)\left(\d x^j+ g^{rj}\d u\right)-2\,\d u\d r-g^{rr}\d u^2.
\end{equation}
Here, $u=\,$const are the null hypersurfaces to which {\boldmath $k$} is normal, $r$ is the affine parameter along the geodesics generated by {\boldmath $k$}${\,=\partial_r}$, and $x \equiv (x^i) \equiv (x^1, x^2, \ldots, x^{D-2})$ are spatial coordinates on a ``transverse'' $(D-2)$-dimensional Riemannian manifold $\M_{(D-2)}$. The metric functions
\begin{equation}\label{cov_contra}
   g^{ri}= g^{ij}g_{uj} , \qquad g^{rr}=-g_{uu}+g^{ij}g_{ui}g_{uj} , \qquad \mbox{and} \qquad g_{ui}= g^{rj}g_{ij} ,
\end{equation}
may depend arbitrarily on $(x,u,r)$, while the spatial components $g_{ij}$ have the factorized form $g_{ij}=p^{-2}(x,u,r)h_{ij}(x,u)$, and $g_{rr}=0=g_{ri}$ (note that $\det g_{ij}=-\det g_{\alpha\beta}$). The expansion of {\boldmath $k$} is given by $\theta\equiv k^\alpha_{\ ;\alpha}/(D-2)=-(\ln p)_{,r}$, which we assume non-vanishing.
The above metric is invariant under the coordinate transformations 
\be
 x^i=x^i(\tilde x,\tilde u) , \qquad u=u(\tilde u), \qquad r=r_0(\tilde x,\tilde u)+\tilde r/\dot u(\tilde u) .
 \label{freedom}
\ee

The next step is to impose Einstein's equations with a suitable energy-momentum tensor in the above Robinson--Trautman class. In the present paper we concentrate on spacetimes with Maxwell fields {\em aligned with the geometrically privileged null vector field} {\boldmath $k$}, characterized by
\begin{equation}
F_{\alpha\beta} k^{\beta}=\eigen \, k_\alpha,
\end{equation}
where $\eigen$ is an arbitrary funtion.
In the coordinate system introduced above this means
\begin{equation}\label{Trivial Implications}
F_{ri}= 0= F^{ui}, \qquad F_{ru}=\eigen=F^{ur},
\end{equation}
with components $F_{ij}, F_{ui}$ (or $F^{ij}=g^{ik}g^{jl}F_{kl}$, $F^{ir}=-\eigen g^{ri}+g^{ij}F_{uj}-g^{rk}g^{ij}F_{kj}$) still arbitrary. Consequently,
\begin{eqnarray}
  {F^u}_r = {F^u}_i= {F^i}_r = & 0 & = {F_r}^u={F_r}^i={F_i}^u, \nonumber\\
  {F^r}_r=-{F^u}_u = & \eigen & = -{F_r}^r = {F_u}^u,
\end{eqnarray}
with ${F^r}_u, {F^r}_i, {F^i}_u, {F^i}_j$, and ${F_i}^r, {F_i}^j, {F_u}^r, {F_u}^i$ generally non-trivial.

For the corresponding energy-momentum tensor of the electromagnetic field
\begin{equation}\label{Energy momentum}
  T_{\alpha\beta} = \frac{1}{4\pi} \left( F_{\alpha \mu} {F_\beta}^\mu- \frac{1}{4} g_{\alpha\beta} F_{\mu\nu} F^{\mu\nu} \right) ,
\end{equation}
we find
\begin{equation}\label{T_rr_T_ri}
  T_{rr} = T_{ri}=0,
\end{equation}
with the remaining components $T_{ij}, T_{ur}, T_{ui}, T_{uu}$ in principle non-trivial and specified below. Notice that the trace %
\begin{equation}
  T_\mu^{\, \mu} = \frac{4-D}{16 \pi} F_{\mu\nu} F^{\mu\nu} = \frac{4-D}{16 \pi} (F_{ij} F^{ij} - 2\eigen^2)
  \label{trace}
\end{equation}
is generally non-zero unless $D=4$.

The field equations $R_{\alpha \beta} - \frac{1}{2} R g_{\alpha \beta} + \Lambda g_{\alpha \beta} = 8 \pi T_{\alpha \beta}$ including an arbitrary {\em cosmological constant}~$\Lambda$ thus take the form 
\begin{equation}\label{Ricci}
R_{\alpha \beta} = \frac{2}{D-2} \Lambda g_{\alpha \beta} + 8 \pi T_{\alpha \beta} + \frac{1}{2} \frac{D-4}{D-2} g_{\alpha \beta} F_{\mu\nu} F^{\mu\nu},
\end{equation}
which will now be solved together with source-free Maxwell equations ${F_{[\alpha\beta;\gamma]}=0}$ and ${{F^{\mu\nu}}_{;\nu} = 0}$, and their Chern--Simons modification in odd dimensions.

\section{Integration of the Einstein--Maxwell field equations}
\label{Einstein}

\subsection{Equations $R_{rr}=0$ and $R_{ri}=0$}
\label{subsec_Rrr}

Due to (\ref{T_rr_T_ri}) and (\ref{geo_metric}), the Einstein equations (\ref{Ricci}) for $R_{rr}$ and $R_{ri}$ are exactly the same as in the vacuum case \cite{PodOrt06}. Consequently, for the Robinson--Trautman class of spacetimes, we obtain $p=r^{-1}$ (up to a trivial rescaling of $h_{ij}$ by a function of $(x,u)$) \cite{PodOrt06}, i.e.
\be
 g_{ij}=r^2h_{ij}(x,u),
 \label{hspatial}
\ee
and
\be
 g^{ri}=e^i(x,u)+r^{1-D}f^i(x,u) ,
  \label{gri}
\ee
where $h_{ij}$, which is the transverse spatial part of the metric, and $e^i, f^i$ are arbitrary functions of $x$ and $u$. The $r$-dependence of the metric functions $g_{ij}, g^{ri}$ is now completely fixed.

Also, thanks to (\ref{hspatial}), we can write
\begin{equation}\label{determinants}
  -\det g_{\alpha\beta}  = r^{2(D-2)} h,
\end{equation}
where
\begin{equation}\label{h}
  h = h(x,u) \equiv \det h_{ij}(x,u).
\end{equation}
We further note that the expansion of the congruence {\boldmath $k$} is now given by $\theta=1/r$.

\subsection{Maxwell equations (step one)}

To determine the $r$-dependence of the components $F_{\mu\nu}$, we now employ Maxwell's equations. With eq.~(\ref{Trivial Implications}), the ``geometrical'' equations ${F_{[\alpha\beta;\gamma]}=0}$, equivalent to ${F_{\alpha\beta,\gamma} + F_{\beta\gamma,\alpha} + F_{\gamma\alpha,\beta}=0}$, imply
\begin{eqnarray}
 F_{ij,r} &=& 0, \label{F_ijr}\\
 F_{ui,r} &=& - \eigen_{,i}, \label{F_uir}\\
 F_{ij,u} &=& F_{uj,i} - F_{ui,j} , \label{F_iju}\\
 F_{[ij,k]}&=& 0. \label{F_ijk} 
\end{eqnarray}
In view of~(\ref{determinants}), the ``dynamical'' equations  ${{F^{\mu\nu}}_{;\nu} = (-\det g_{\alpha\beta})^{-\frac12}\big((-\det g_{\alpha\beta})^\frac12  F^{\mu\nu}\big)_{,\nu}=0}$ are
\begin{eqnarray}
 (r^{D-2} \eigen)_{,r} &=& 0, \label{eigen_r}\\
 \sqrt{h} \;(r^{D-2} \;  F^{ir})_{,r} &=& -r^{D-2} \; (\sqrt{h} \; F^{ij})_{,j}, \label{eigen_j}\\
 (\sqrt{h} \; F^{ir})_{,i} &=& -(\sqrt{h} \; \eigen)_{,u}. \label{eigen_u}
\end{eqnarray}

From (\ref{F_ijr}) we observe that the components $F_{ij}$ are independent of $r$,
\begin{equation}\label{Fij}
F_{ij} = F_{ij} (x,u).
\end{equation}
Using (\ref{eigen_r}), we find
\begin{equation}\label{Fru}
  F_{ru} =\eigen= r^{2-D} \q(x,u),
\end{equation}
with $\q(x,u)$ arbitrary. Using this result and (\ref{F_uir}), we obtain
\begin{equation}
F_{ui}= r^{3-D} \frac{\q_{,i}}{D-3} - \xi_i (x,u), \label{Fui}
\end{equation}
with $\xi_i(x,u)$ being some functions of $x$ and $u$. Thus we found the $r$-dependence of all electromagnetic field components. In particular, the invariant $F_{\mu \nu} F^{\mu\nu}$ of the Maxwell field is
\begin{equation}
  F_{\mu \nu} F^{\mu\nu} = r^{-4} \etta^2 - r^{2(2-D)}\, 2 \q^2, \label{F_invariant}
\end{equation}
where we have defined
\begin{equation}
  \etta^2(x,u) \equiv F_{ik} F_{jl} h^{ij} h^{kl} ,\label{eta2}
\end{equation}
and (from now on) $h^{ij}$ denotes the inverse of $h_{ij}$.
We always have $\etta^2 \ge 0$ (with $\etta^2=0 \Leftrightarrow F_{ij}=0$) because, in an orthonormal frame, $\etta^2=\sum_{i,j}F_{(i)(j)}^2$. 

Substituting (\ref{Fui}) into  (\ref{F_iju}), we get
\begin{eqnarray}
&& F_{ij,u} =  \xi_{i,j} - \xi_{j,i}\> , \label{F_ijuN}
\end{eqnarray}
while equation (\ref{F_ijk}) is unchanged. Finally, if expanded in the powers of $r$ using the previous results,
the remaining Maxwell equations (\ref{eigen_j}) and (\ref{eigen_u}) yield the following set of relations in ${D>4}$:
\begin{eqnarray}
 \q f^i &=& 0, \label{c1}\\
 F_{jk}f^k   &=& 0, \label{c2}\\
 \q_{,j} &=& 0, \label{c3}\\
 \xi_j-F_{jk}e^k &=& 0, \label{c4}\\
 (\sqrt{h}\,h^{ik}h^{jl}F_{kl})_{,j} &=& 0, \label{c5}\\
 (\sqrt{h}\,\q)_{,u}-(\sqrt{h}\,\q\, e^i)_{,i} &=& 0. \label{c6}
\end{eqnarray}
Relations (\ref{F_ijuN})--(\ref{c6}) and (\ref{F_ijk}) place restrictions on the admissible electromagnetic fields (\ref{Fij})--(\ref{Fui}) in Robinson--Trautman spacetimes. We shall return to the implications of these constraints after we employ the following Einstein equation in subsection~\ref{subsec_Rij}. For the special case ${D=4}$, see the Appendix.

The above results have already an important consequence. Namely, one of the necessary conditions for having a {\em null Maxwell field} reads $F_{\mu\nu} F^{\mu\nu}=0$. In view of the $r$-dependence specified by~(\ref{F_invariant}), one finds immediately that for $D>4$ this requires $F_{ij}=0=\q$ and thus $F_{ru}=0$. Substituting into~(\ref{eigen_j}) this gives also $F_{ui}=0$, that is, a vanishing electromagnetic field. Hence {\em higher dimensional Robinson--Trautman spacetimes do not admit aligned null Maxwell fields}, as opposed to the $D=4$ case \cite{RobTra62,Stephanibook}. This is an explicit example of the result of \cite{Ortaggio07} that higher dimensional null Maxwell fields can not have expanding rays with vanishing shear.

\subsection{Chern--Simons term}
\label{subsec_CS}

In theories which include a Chern--Simons term, formulated in  odd spacetime dimensions (${D=2n+1}$), the set of geometrical equations (\ref{F_ijr})--(\ref{F_ijk}) is unchanged ($\d {\mbox{\boldmath$F$}}=0$). 
On the other hand, the dynamical set contains now an additional term on the r.h.s. (cf., e.g., \cite{GauMyeTow99})
\be
 (\sqrt{-\det g_{\alpha\beta}} \; F^{\mu\nu})_{,\nu} =-\lambda\,\epsilon^{\mu\gamma\delta\ldots\sigma\tau}\underbrace{F_{\gamma\delta}\ldots F_{\sigma\tau}}_{\mbox{ \tiny $n$ times}} ,
 \label{CS}
\ee
where $\lambda$ is a coupling constant. Note that $F_{ri}=0$ thanks to the alignment condition~(\ref{Trivial Implications}), so that the Chern--Simons term does not affect eq.~(\ref{CS}) with $\mu=u$, which thus takes again the form~(\ref{eigen_r}). In fact, this is the only dynamical equation we used in the discussion above, so that eqs. (\ref{Fij})--(\ref{F_ijuN}) apply also in the Chern--Simons case. Moreover, if one assumes that also $F_{ij}=0$ (no ``magnetic'' field), the Chern--Simons term then vanishes identically for any odd $D$. In particular, null fields are thus  ruled out again. Further analysis will be simpler after looking at the next Einstein equation.

\subsection{Equation $R_{ij}=\frac{2}{D-2}\Lambda g_{ij} + 8 \pi T_{ij} + \frac{1}{2} \frac{D-4}{D-2} g_{ij}F_{\mu\nu} F^{\mu\nu} $}
\label{subsec_Rij}

The Ricci tensor component $R_{ij}$ for the metric (\ref{geo_metric}) was calculated in \cite{PodOrt06}. With (\ref{hspatial}), (\ref{gri}) this reads
\begin{eqnarray}
 && R_{ij}\>=\> {\mathcal R_{ij}}-r^{4-D}\left(r^{D-3}g^{rr}\right)_{,r}h_{ij}-r^{2(2-D)}\frac{(D-1)^2}{2}h_{ik}h_{jl}  f^{k}f^{l} \nonumber\\
 && {\qquad}-r\left[\frac{D-2}{2}\Big(2h_{k(i}{e^{k}}_{,j)}+e^{k}h_{ij,k}-h_{ij,u}\Big)
    +\left(e^{k}_{\;,k}+e^{k}(\ln\sqrt{h})_{,k}-(\ln\sqrt{h})_{,u}\right)h_{ij}\right]\nonumber\\
 && {\qquad}+r^{2-D}\left[\frac{1}{2}\left(2h_{k(i}{f^{k}}_{,j)}+f^{k}h_{ij,k}\right)
    -\left(f^{k}_{\;,k}+f^{k}(\ln\sqrt{h})_{,k}\right)h_{ij}\right] , 
\end{eqnarray}
where
${\mathcal R_{ij}}$ is the Ricci tensor associated with the spatial metric $h_{ij}$, and indices in small round brackets are symmetrized. The corresponding component of the energy-momentum tensor is
\begin{equation}\label{Tij}
T_{ij} = \frac{1}{8 \pi} r^{2(3-D)} \q^2 h_{ij} + r^{-2} \frac{1}{4 \pi} \Big( F_{ik} F_{jl} h^{kl} - \frac{1}{4} \etta^2 h_{ij} \Big).
\end{equation}
Using (\ref{Fij}) and (\ref{F_invariant}), one can separate terms in the field equation with different $r$-dependence. By contracting with $h^{ij}$, we obtain a differential equation for $g^{rr}$ which can be integrated immediately. For $D>5$ this yields
\be
 g^{rr}=c_1+c_2r+c_3r^2+c_4r^{2-D}+c_5r^{2(2-D)}+c_6r^{3-D} + c_7 r^{-2} + c_8 r^{2(3-D)},
 \label{grr_old}
\ee
where $c_1,\ldots, c_8$ are functions of $(x,u)$ as follows
\beqn
 & & c_1=\frac{{\cal R}}{(D-2)(D-3)}\, , \qquad\quad c_2=\frac{2}{D-2}\left[(\ln\sqrt{h})_{,u}-e^k_{\;,k}-e^k (\ln\sqrt{h})_{,k}\right] ,   \nonumber\\
 & & c_3=-\frac{2\Lambda}{(D-1)(D-2)}\, , \qquad c_4=\frac{D-3}{D-2}\left[f^k_{\;,k}+f^k (\ln\sqrt{h})_{,k}\right] , \nonumber\\
 & & c_5=\frac{1}{2}\frac{D-1}{D-2}h_{kl}f^k f^l\, , \qquad \quad c_6 \ \> \mbox{arbitrary}\, ,  \label{c_i}\\
 & & c_7=-\frac{\etta^2}{(D-2)(D-5)}\, , \qquad c_8= \frac{2\q^2}{(D-2)(D-3)}\, , \nonumber
\eeqn
where ${\cal R}=h^{ij}{\mathcal R_{ij}}$. For $D=5$ the only difference is that in (\ref{grr_old}) one should replace $c_7 r^{-2}$ with the term $-\frac{1}{3}\etta^2r^{-2}\ln(c_7r)$ where $c_7$ is an arbitrary function of $(x,u)$ with the dimension of an inverse length.
Next, substituting the above expressions back into the Einstein equation for $R_{ij}$, we determine for \emph{any} ${D>4}$ the following constraints on the metric $h_{ij}$ and the functions $e^i$ and $f^i$: 
\beqn
 {\mathcal R}_{ij}&=&\frac{{\mathcal R}}{D-2}h_{ij} , \label{constr_c1}\\
2h_{k(i}{e^k }_{,j)} +e^k h_{ij,k}-h_{ij,u}&=&\frac{2}{D-2}\left[e^k_{\;,k}+e^k (\ln\sqrt{h})_{,k} -(\ln\sqrt{h})_{,u}\right]h_{ij} , \label{constr_c2} \\
 2h_{k(i}{f^k }_{,j)} +f^k h_{ij,k}&=&\frac{2}{D-2}\left[f^k_{\;,k}+f^k (\ln\sqrt{h})_{,k}\right]h_{ij}, \label{constr_c4}\\
(h_{kl}f^k f^l)\,h_{ij} &=& (D-2) (h_{ik}f^k)(h_{jl}f^l) \label{constr_c5},\\
h_{ij} \etta^2 & = & (D-2) F_{ik} F_{jl} h^{kl} \label{constr_c7}.
\eeqn

As we notice, (\ref{constr_c5}) is identical to the vacuum case discussed in \cite{PodOrt06} and it requires
\be
 f^i=0 .
 \label{f^i}
\ee
In analogy to \cite{PodOrt06}, we also use the coordinate freedom~(\ref{freedom}) to achieve
\be
 e^i=0,
 \label{e^i}
\ee
so that $g^{ri}=0=g_{ui}$ and eqs.~(\ref{constr_c4}) and (\ref{constr_c5}) are now  satisfied identically.

In addition, constraint (\ref{constr_c7}) requires 
${\etta^2=0}$ for \emph{any odd} $D$. Indeed, taking the determinant of (\ref{constr_c7}), we obtain $(\etta^2)^{D-2} h^2 = (2-D)^{D-2} (\det F_{ij})^2$, but ${\,\det F_{ij}=0\,}$ for any antisymmetric matrix~$F_{ij}$ and odd dimension $D-2$ since $\det F_{ij} = \det (-F_{ij}) = (-1)^{D-2} \det F_{ij}$. Consequently, the logarithmic term in ${D=5}$ is zero, and in the following we need not treat the ${D=5}$ case separately.

The Robinson--Trautman metric is thus simplified considerably and reads
\be
 \d s^2=r^2h_{ij}\,\d x^i\d x^j-2\,\d u\d r-g^{rr}\d u^2 ,
 \label{geo_metric spec}
\ee
where, using eqs.~(\ref{grr_old}) and (\ref{c_i}), the coefficient $g^{rr}$ is explicitly given by
\begin{eqnarray}
g^{rr} & = & \frac{{\cal R}}{(D-2)(D-3)}+\frac{2(\ln\sqrt{h})_{,u}}{D-2}\,r-\frac{2\Lambda}{(D-2)(D-1)}\,r^2-\frac{\mu}{r^{D-3}} \nonumber\\
&& {}+ \frac{2 \q^2}{(D-2)(D-3)} \frac{1}{r^{2(D-3)}}- \frac{\etta^2}{(D-2)(D-5)} \frac{1}{r^{2}} \, , \label{grr}
\end{eqnarray}
with
\be
\etta^2=0\qquad \mbox{for any odd\ }  D= 5, 7, 9, \ldots .
\ee
The function $\mu(x,u)$, which renames~$c_6$, is arbitrary. 

The $(D-2)$-dimensional spatial metric $h_{ij}$
is constrained by (\ref{constr_c1}) and (\ref{constr_c2}), now with $e^i=0$:
\beqn
 {\mathcal R}_{ij}&=&\frac{{\mathcal R}}{D-2}h_{ij} , \label{constrijr} \\
  h_{ij,u}&=&\frac{2}{D-2} h_{ij}(\ln\sqrt{h})_{,u} . \label{constrijs}
\eeqn
As in \cite{PodOrt06}, relation (\ref{constrijr}) tells us that at any given $u=u_0=\,$const, the spatial metric $h_{ij}(x,u_0)$ must describe an Einstein space $(\M_{(D-2)},h_{ij})$. 
For ${D>4}$ this implies \cite{PodOrt06} that the spatial Ricci scalar ${\mathcal R}$ can \emph{only} depend  on the coordinate $u$ (and that in the particular case $D=5$ the metric $h_{ij}(x,u_0)$ corresponds to a 3-space of constant curvature).
Equation (\ref{constrijs}) ``controls'' the parametric dependence of $h_{ij}(x,u)$ on $u$, and can easily be integrated to obtain $h_{ij}=h^{1/(D-2)}\,\gamma_{ij}(x)$. Consequently, $ h\equiv \det h_{ij}=h \det \gamma_{ij}$, so that the matrix $\gamma_{ij}$ must be unimodular. Considering eq.~(\ref{h}) we can write
\be
h_{ij}=\frac{\gamma_{ij}(x)}{P^2(x,u)} \qquad\hbox{where}\qquad \det \gamma_{ij}=1 , \quad P^{-2}=h^{1/(D-2)} . \label{hijs}
\ee
The spatial metric $h_{ij}(x,u)$ can thus depend on the coordinate $u$ only via the conformal factor $P^{-2}$.

\subsection{Maxwell equations (step two)}

\label{subsec_Max2}

Let us now return to the Maxwell equations. As noticed above, for any odd $D$ we have $\etta^2 = 0$, i.e.
\begin{equation} \label{Zero Maxwell Components}
  F_{ij}=0 \qquad (D=2n+1 \mbox{ odd}) .
\end{equation}
Thanks to this significant simplification in odd dimensions, {\em the Chern--Simons term in eq.}~(\ref{CS}) {\em vanishes identically} (cf. the discussion in subsection~\ref{subsec_CS}), and from now on we can thus study both Maxwell and Maxwell--Chern--Simons theories in a unified way.

Since now $e^i=0=f^i$ (cf.~(\ref{f^i}), (\ref{e^i})), the dynamical Maxwell equations (\ref{c1})--(\ref{c4}) simplify to $\q_{,j}=0$, $\xi_j=0$. In view of (\ref{Fui}),\, (\ref{Fru}), and (\ref{F_ijuN}), we see that for $D>4$
\begin{equation}\label{Maxwell field}
  F_{ui}=0, \qquad F_{ru} = \frac{\q(u)}{r^{D-2}}, \qquad F_{ij}=F_{ij}(x).
\end{equation}

The only remaining Maxwell equations (\ref{F_ijk}), (\ref{c5}) and (\ref{c6}) read
\begin{eqnarray}
 F_{[ij,k]} &=& 0, \label{constrijF1}\\
 (\sqrt{h}\,h^{ik}h^{jl}F_{kl})_{,j} &=& 0, \label{nc5}\\
 (\sqrt{h}\,\q)_{,u} &=& 0. \label{nc6}
\end{eqnarray}%
Notice that in even dimensions (cf.~(\ref{Zero Maxwell Components})) relations (\ref{constrijF1}) and (\ref{nc5}) are effective source-free Maxwell equations for the $(D-2)$-dimensional ``spatial'' (magnetic) field $F_{ij}$ \emph{in the Riemannian geometry of}~$h_{ij}$.  That is, the 2-form
\begin{equation}\label{spacefield}
  \mbox{\boldmath$\tilde F$}\equiv{\textstyle\frac{1}{2}}F_{ij}(x) \, \d x^i \wedge \d x^j
\end{equation}
must be closed ($\d \mbox{\boldmath$\tilde F$}=0$) and coclosed ($\d{^*}\mbox{\boldmath$\tilde F$}=0$) in $(\M_{(D-2)},h_{ij})$. However, $\mbox{\boldmath$\tilde F$}$ must also obey the extra constraint~(\ref{constr_c7}), that is the last remnant of the Einstein equation for $R_{ij}$. 
Recalling the block-diagonal canonical form 
$\mbox{\boldmath$\tilde F$}=c_{12}\,\mbox{\boldmath$m$}^{(1)}\wedge\mbox{\boldmath$m$}^{(2)}+c_{34}\,\mbox{\boldmath$m$}^{(3)}\wedge\mbox{\boldmath$m$}^{(4)}+\ldots$
of a generic even-dimensional antisymmetric matrix in an adapted orthonormal coframe $(\mbox{\boldmath$m$}^{(1)},\ldots,\mbox{\boldmath$m$}^{(D-2)})$ of $h_{ij}$, the condition~(\ref{constr_c7}) requires that in such a coframe one has in fact
\be \mbox{\boldmath$\tilde F$}=\frac{\etta}{\sqrt{D-2}}\left(\mbox{\boldmath$m$}^{(1)}\wedge\mbox{\boldmath$m$}^{(2)}+\mbox{\boldmath$m$}^{(3)}\wedge\mbox{\boldmath$m$}^{(4)}+\ldots+\mbox{\boldmath$m$}^{(D-3)}\wedge\mbox{\boldmath$m$}^{(D-2)}\right) \qquad (D=2n+2 \mbox{ even}). \label{Max_spatial}
\ee

This special form of $\mbox{\boldmath$\tilde F$}$ implies 
\be 
 {^*}\mbox{\boldmath$\tilde F$}=\frac{(2n)^{(n-2)/2}}{(n-1)!}\etta^{-(n-2)}\underbrace{\mbox{\boldmath$\tilde F$}\wedge \mbox{\boldmath$\tilde F$}\wedge\ldots\wedge \mbox{\boldmath$\tilde F$}}_{\mbox{ \tiny $(n-1)$ times}} \ ,
 \label{spec_Max}
\ee
where the ${^*}$-duality and $\wedge$-product are (in this paragraph only) those of $(\M_{(D-2)},h_{ij})$. 
Hence, for $n>2$ ($D>6$) imposing that the 2-form $\mbox{\boldmath$\tilde F$}$ is simultaneosly closed and coclosed requires $\etta_{,i}=0$. For $n=2$ ($D=6$), instead, $\mbox{\boldmath$\tilde F$}$ is self-dual, therefore if it is closed it is also automatically coclosed, without any restriction on $\etta$. We will recover the same results explicitly also below using the Einstein equations, cf.~eq.~(\ref{eta2i}).

Note also that if $\mbox{\boldmath$\tilde F$}$ is supposed to be regular and non-zero on $\M_{(D-2)}$, then eq.~(\ref{constr_c7}) requires that the Einstein space $(\M_{(D-2)},h_{ij})$ is an almost-Hermitian (possibly, Hermitian) manifold \cite{KobNom2} with the almost-complex structure $J^i_{\, j}=|\etta|^{-1}(D-2)^{1/2}F^i_{\ j}$. In view of the previous comments, for $D=2n+2>6$ the Maxwell equations imply that the 2-form $J_{ij}=h_{ik}J^k_{\, j}$ associated with the almost-complex structure is closed, so that the transverse space is not only almost-Hermitian but actually almost-K\" ahler (possibly, K\" ahler).

\subsection{Equation $R_{ur}=-\frac{2}{D-2}\Lambda + 8 \pi T_{ur} - \frac{1}{2} \frac{D-4}{D-2} F_{\mu\nu} F^{\mu\nu}$}
The Ricci tensor component $R_{ur}$ for the metric (\ref{geo_metric spec}) reads
$R_{ur}= \frac{1}{2}r^{2-D}\left(r^{D-2}g^{rr}_{\;\;\; ,r}\right)_{,r} -r^{-1}(\ln\sqrt{h})_{,u}$, see \cite{PodOrt06}.
By substituting the expression (\ref{grr}) we obtain
\beqn
&& R_{ur}= -\frac{2\Lambda}{D-2} + r^{2(2-D)} \frac{D-3}{D-2}\, 2\q^2 + r^{-4} \frac{1}{D-2}\,\etta^2  \,  .
\eeqn
Using (\ref{F_invariant}) and
\begin{equation}
T_{ur} = \frac{1}{16 \pi} F_{\mu \nu} F^{\mu\nu}+ \frac{1}{4 \pi} r^{2(2-D)} \,\q^2,\label{Tur}
\end{equation}
we observe that the corresponding field equation is automatically satisfied in any dimension.

\subsection{Equation $R_{ui}= 8 \pi T_{ui}$}
For the energy-momentum tensor, using (\ref{Energy momentum}) and (\ref{Maxwell field}), we find $T_{ui} = 0$. The Ricci tensor component $R_{ui}$ for the metric (\ref{geo_metric spec}) and (\ref{grr}), using $\q_{,i}=0$ and relation (\ref{constrijs}), is
\begin{equation}
 R_{ui} = r^{-1} \frac{{(D-4) \cal R}_{,i}}{2(D-2)(D-3)} + r^{2-D} \; \frac{\mu_{,i}}{2} - r^{-3} \; \frac{(D-6)(\etta^2)_{,i}}{2(D-2)(D-5)}.
\end{equation}
Comparing the coefficients of different powers of $r$, we obtain immediately the following conditions
\begin{eqnarray}
  (D-4)\,{\cal R}_{,i} & = & 0, \label{Ri}\\
  \mu_{,i} & = & 0, \label{mui} \\
  (D-6)(\etta^2)_{,i} & = & 0. \label{eta2i}
\end{eqnarray}
Therefore, (for $D>4$) the functions $\cal{R}$ and $\mu$ must be independent of the spatial coordinates,
\be \label{no spatial dependance}
{\cal R}={\cal R}(u), \qquad \mu=\mu(u),
\ee
and we further find
\begin{equation}
  \etta^2=\etta^2 (u) \qquad \mbox{for} \qquad D \not= 6,  \mbox{ even},
  \label{not6}
\end{equation}
with $\etta^2=0$ in any odd $D$. For $D=6$, eq.~(\ref{eta2i}) is satisfied identically, with $\etta^2$ remaining a function of both $x$ and $u$. In $D=4$, corresponding to the standard General Relativity, equation (\ref{Ri}) is an identity, so that one can have a much more general function ${\cal R}(x,u)$; eqs.~(\ref{mui}) and (\ref{eta2i}) are also modified, cf.~the Appendix.

\subsection{Equation $R_{uu}=\frac{2}{D-2}\Lambda g_{uu}+ 8 \pi T_{uu} + \frac{1}{2} \frac{D-4}{D-2} F_{\mu \nu} F^{\mu \nu} g_{uu}$}
\label{subsec_Ruu}

Finally, we evaluate the Ricci tensor component $R_{uu}$. Using the general expression (31) of \cite{PodOrt06}, relation (\ref{grr}) for $g^{rr}=-g_{uu}$, equations (\ref{constrijs}), (\ref{hijs}) implying $\sqrt h=P^{2-D}$, and (\ref{no spatial dependance}), we obtain
\beqn
&& R_{uu} = \frac{2}{D-2}\Lambda g_{uu} - \left( r^{-4} \frac{\etta^2}{D-2} + r^{2(2-D)} \frac{D-3}{D-2}\, 2\q^2 \right) g_{uu}\nonumber\\
&&  {}+ r^{2-D} \frac{D-2}{2}\,\left[ (D-1)\mu(\ln P)_{,u}-\mu_{,u} \right] - r^{5-2D} \frac{2\q}{D-3} \left[ (D-2)\q(\ln P)_{,u}-\q_{,u} \right] \nonumber \label{Ruu} \\
&&  {}+ \frac{r^{-3}}{2(D-5)}\,\left[ 4(\etta^2)(\ln P)_{,u}-(\etta^2)_{,u} \right] - \frac{r^{-4}\Delta(\etta^2)}{2(D-2)(D-5)}.
\eeqn
where $\Delta$ is the covariant Laplace operator with respect to the spatial metric $h_{ij}$, i.e. $\Delta(\etta^2) \equiv (\etta^2)^{\|j}_{\>\|j} = \left[ (\etta^2)_{,i} h^{ij}\right]_{,j} + (2-D) h^{ij} (\etta^2)_{,i} (\ln P)_{,j}$ . Note that we also dropped the term proportional to $r^{-1}$, which vanishes identically, see \cite{PodOrt06} (eqs.~(33) and (B.4) therein). 

Now, the coefficient of the $r^{5-2D}$ term vanishes provided the Maxwell equation~(\ref{nc6}) is satisfied, and the coefficient of $r^{-3}$ is zero thanks to $F_{ij}=F_{ij}(x)$ and eqs.~(\ref{eta2}) and (\ref{hijs}) --- indeed these conditions can be reexpressed as
\begin{eqnarray}
    \q_{,u} & = & (D-2) (\ln P)_{,u} \; \q \  , \label{id_phi} \\
   (\etta^2)_{,u} & = & 4 (\ln P)_{,u} \; (\etta^2)\ .\label{logarithmic derivative of eta}
\end{eqnarray}
Moreover, using (\ref{F_invariant}) and (\ref{Maxwell field}), we have
\begin{equation}
8 \pi T_{uu} + \frac{1}{2} \frac{D-4}{D-2} F_{\mu \nu} F^{\mu \nu} g_{uu} = - \left( r^{-4} \frac{\etta^2}{D-2} + r^{2(2-D)} \frac{D-3}{D-2} 2\q^2 \right) g_{uu}.
\end{equation}
We thus now only need to make sure that in~(\ref{Ruu}) the coefficients of $r^{2-D}$ and of the last term in $r^{-4}$ vanish. Note that, by~(\ref{not6}), $(\etta^2)_{,i}=0$ for $D\neq 6$ so that the latter is automatically zero. On the other hand, in the special case $D=6$ both terms are non-zero and they combine in a single expression. The field equations thus require
\be
 \mu_{,u}=(D-1) (\ln P)_{,u} \; \mu \, \quad\qquad(D\neq 4,6) \label{logarithmic derivative of mu} ,
\ee 
or 
\be
\mu_{,u}  =5(\ln P)_{,u}\;\mu-{\textstyle\frac{1}{16}}\Delta(\etta^2) \qquad (D=6) \label{special6} ,
\ee 
in the two distinct cases $D\neq 6$ and $D=6$, which we analyze separately in the next section.

\section{Summary and discussion}
\label{sec_discussion}

Starting from the general Robinson--Trautman geometric ansatz~(\ref{geo_metric}), in the preceding section we have imposed all the constraints coming from the Einstein--Maxwell equations. The resulting metric takes a simplified form~(\ref{geo_metric spec}), which is fully specified by the single function $g^{rr}$ in eq.~(\ref{grr}), along with the transverse Einstein geometry $(\M_{(D-2)},h_{ij})$, as determined by (\ref{constrijr}). The specific form of the parameters and functions entering eq.~(\ref{grr}) and possible constraints on the Einstein metrics $h_{ij}$ depend on the number of spacetime dimensions, as we will now discuss in the following.

\subsection{Even dimensions: the generic case ($D\neq 6$)}

\label{subsec_evengen}

For an arbitrary even ${D>4}$ such that ${D\neq 6}$, by differentiating any of eqs.~(\ref{id_phi}), (\ref{logarithmic derivative of eta}), (\ref{logarithmic derivative of mu}) with respect to the spatial coordinates, we obtain (recall that $\q=\q(u)$, $\etta=\etta(u)$, $\mu=\mu(u)$)
\begin{equation}
  (\ln P)_{,ui} = 0  , 
  \label{P_sep}
\end{equation}
unless $\mu =0$ and $\q =0 = \etta^2$, which is the exceptional vacuum spacetime discussed in \cite{PodOrt06,Ortaggio07}, and \cite{PraPraOrt07}. Eq.~(\ref{P_sep}) can be integrated immediately, yielding the factorized form ${ P(x,u) = P(x) \, U(u)}$,
where $P$ and $U$ are arbitrary functions. Without loss of generality, we can set $U=1$ by a suitable coordinate transformation of the form $u=u(\tilde u),\   r=\tilde r/\dot u(\tilde u)$, under which the form of the metric (\ref{geo_metric spec}), (\ref{grr}) is invariant and the individual metric functions are reparameterized as follows
\begin{equation}
  \tilde P=P\,\dot u\,,\ \tilde{\cal R}={\cal R}\,{\dot u}^2\,,\ \tilde \mu=\mu\,{\dot u}^{D-1}\,, \ \tilde \etta^2=\etta^2\,{\dot u}^{4}\,, \ \tilde \q=\q\,{\dot u}^{D-2}.
\label{scaling}
\end{equation} 
Choosing $\dot{u}=1/U$ and dropping tildas, we obviously achieve
\begin{equation}
  P(x,u) = P(x),
  \label{factorizedP}
\end{equation}
and considering eqs.~(\ref{id_phi}), (\ref{logarithmic derivative of eta}), and (\ref{logarithmic derivative of mu}), we thus have
\begin{equation}
  \mu = \mbox{const}, \qquad \q = \mbox{const}, \qquad \etta^2= \mbox{const}.
  \label{constants}
\end{equation}
Considering (\ref{no spatial dependance}) and the fact that ${\cal R}$ is the Ricci scalar associated with the spatial metric $h_{ij}=h_{ij}(x)=P^{-2}(x) \gamma_{ij}(x)$, cf.~(\ref{hijs}), which now does not involve $u$, we conclude
\begin{equation}
  \cal R = \mbox{const}.
\end{equation}
In addition, we can now always set the constant term $K\equiv{\cal R}/(D-2)(D-3)$ in the metric (\ref{grr}) to ${K = \pm 1, 0}$ using the remaining  scaling freedom (\ref{scaling}), namely $u \rightarrow {\cal C} u, r \rightarrow r/{\cal C}, x^i \rightarrow {\cal C}^{1/(D-2)} x^i$. 

To summarize, the explicit form of even dimensional ($D\neq 6$) Robinson--Trautman spacetimes with an aligned electromagnetic field and possibly a cosmological constant is
\be
 \d s^2=r^2\,h_{ij}(x)\,\d x^i\d x^j-2\,\d u\d r-2H(r)\,\d u^2 .
 \label{geo_metric fin}
\ee
The function $2H\equiv g^{rr}=-g_{uu}$ and the Maxwell field are given by
\begin{equation}\label{general form of H}
2H = K-\frac{2\Lambda}{(D-2)(D-1)}\,r^2-\frac{\mu}{r^{D-3}} + \frac{2 \q^2}{(D-2)(D-3)} \frac{1}{r^{2(D-3)}} - \frac{\etta^2}{(D-2)(D-5)} \frac{1}{r^{2}} \,,
\end{equation}
\begin{eqnarray}
  \mbox{\boldmath$F$} = \frac{\q}{r^{D-2}} \d r \wedge \d u + {\textstyle\frac{1}{2}}F_{ij}(x) \, \d x^i \wedge \d x^j \qquad (D\neq 6, \mbox{ even}) , \label{Max_even}
\end{eqnarray}
where ${K=\pm1, 0}$, and ${\mu, \q, \etta}$ are constants.\footnote{In the special case ${D=4}$ (see also the Appendix) the electric and magnetic monopole terms in $\q$ and $\etta$ become indistinguishable in the metric. This corresponds to the well-known fact that in ${D=4}$ Einstein--Maxwell gravity all solutions are determined only up to a costant duality rotation of the electromagnetic field.\label{footn_duality}} The transverse manifold $(\M_{(D-2)},h_{ij})$ is a \emph{Riemannian Einstein space}, see~(\ref{constrijr}), with the Ricci scalar normalized as ${{\cal R}=K(D-2)(D-3)}$. If this is taken to be compact, these solutions admit a \emph{black hole interpretation}, with a static exterior region at $2H>0$. Obviously, $\Lambda$ is the cosmological constant, $\mu$ parametrizes the mass, and $\q$ is the electric charge. If the magnetic term $F_{ij}$ is non-zero, $(\M_{(D-2)},h_{ij},J^i_{\, j})$ must be an {\em almost-K\" ahler Einstein} manifold (cf.~subsection~\ref{subsec_Max2}). The almost-complex structure gives $F_{ij}$ (up to a constant factor), which thus satisfies the ``effective'' $(D-2)$-dimensional Maxwell equations~(\ref{constrijF1}) and (\ref{nc5}). 

Note that, when $\etta$ is non-zero and ${D>4}$, $(\M_{(D-2)},h_{ij})$ can not be a sphere of constant curvature,\footnote{More generally, it is an old result that K\" ahler manifolds of constant (Riemannian) curvature must be flat in $2n>2$ real dimensions \cite{Bochner47}. It has been demonstrated more recently that this applies also to almost-K\" ahler manifolds (see \cite{Armstrong02} and references therein). In this context, it is also worth mentioning that the celebrated conjecture of \cite{Goldberg69} that almost-K\" ahler, Einstein, compact manifolds must be K\" ahler has been proven in the case of non-negative scalar curvature \cite{Sekigawa87}. See, e.g., \cite{Armstrong02,ApoDra03} for some more general properties of almost-K\" ahler Einstein manifolds and for more references.}
as one would require, e.g., for an asymptotically flat spacetime. By constrast, sperically symmetric magnetic monopole solutions of the Einstein--Yang--Mills equations have been recently found in \cite{GibTow06}. The line element given in \cite{GibTow06} coincides with our eqs.~(\ref{geo_metric fin}), (\ref{general form of H}) in the special subcase $K=1$, $\q=0$, except that $(\M_{(D-2)},h_{ij})$ is a round sphere there. Note, however, that even in that case the large-$r$ behaviour of the $\etta$ term in~(\ref{general form of H}) does spoil the standard ``good properties'' of an asymptotically simple spacetime \cite{GibTow06}.  We refer to  \cite{GibTow06} for a discussion of the horizon structure (cf.~also \cite{KodIsh04} for the case $\etta=0$).

From the above form~(\ref{Max_even}) of the Maxwell field, it is clear that it is of type D \cite{Coleyetal04vsi} with principal null directions given by 
\be
 \mbox{\boldmath$k$}=\pa_r , \qquad \mbox{\boldmath$l$}=\pa_u-H\pa_r\ .
 \label{nulldirections}
\ee
In addition, in view of~(\ref{constants}) the line element~(\ref{geo_metric fin}) and~(\ref{general form of H}) is a warped product of $(\M_{(D-2)},h_{ij})$ with a two-dimensional Lorentzian factor. For such type of warped spacetimes, the Weyl tensor is necessarily of type D, unless zero (type O) \cite{PraPraOrt07}. However, the latter case can not occur here since, e.g., the Weyl component $C_{ruru}$ reads 
\be
 C_{ruru}=-(D-2)(D-3)\frac{\mu}{2r^{D-1}}+ \frac{(2D-5)}{(D-1)} \frac{2 \q^2}{r^{2(D-2)}} -  \frac{(D-3)}{(D-1)(D-2)(D-5)}\frac{6 \etta^2}{r^4} \ .
 \label{weylnulldirections} 
\ee
The above Robinson--Trautman spacetimes in ${D>4}$ are thus of type D with WANDs given again by (\ref{nulldirections}) (cf.~\cite{PraPraOrt07}). Conformal flatness requires ${\mu=0}$ and ${\q=0=\etta}$, in particular vacuum spacetimes  \cite{PodOrt06}, so that the only possible conformally flat metrics are of constant curvature.

Note finally that when $F_{ij}=0$ (i.e., $\etta=0$), these solutions are \emph{electrically charged black holes}. In the simplest case when  $(\M_{(D-2)},h_{ij})$ is a round sphere, one obtains the well-known asymptotically flat/(A)dS spacetimes of \cite{Tangherlini63}. However, $(\M_{(D-2)},h_{ij})$ can now be {\em any} Einstein space (cf. also \cite{GibWil87}, and see, e.g., \cite{Birmingham99,GibHar02} for related discussions in the vacuum case ${\q=0=\etta}$). Stability properties of these black holes have been studied in \cite{KodIsh04}.

\subsubsection{An explicit example}

\label{subsubsec_example}

For the sake of definiteness, as a simple example with $\etta\neq 0$ we can consider $(\M_{(D-2)},h_{ij})$ as the Riemannian analog of Nariai-like solutions with geometries ${S^2\times S^2\times\ldots\ }$ or ${\ H^2\times H^2\times\ldots}$, namely
\beqn
 & & h_{ij}\ \d x^i\d x^j=\sum_{I=1}^n\left[\left(1-\epsilon\frac{\rho_I^2}{a^2}\right)\d\psi_I^2+\left(1-\epsilon\frac{\rho_I^2}{a^2}\right)^{-1}\d\rho_I^2\right] , \nonumber \\
 & & {\textstyle\frac{1}{2}}F_{ij}\ \d x^i\wedge\d x^j=\frac{\etta}{\sqrt{D-2}}\sum_{I=1}^n\d\psi_I\wedge\d\rho_I , \qquad (D=2n+2)
\eeqn
where ${\epsilon=+1}$ or ${\epsilon=-1}$ (or ${\epsilon=0}$, which gives a flat $h_{ij}$), $a$ and $\etta$ are constants, the scalar curvature is given by ${K=\epsilon a^{-2}(2n-1)^{-1}}$ (normalizable to $K=\epsilon$ if desired) and ${D=2n+2}$ is the number of spacetime dimensions. Note that $F_{ij}$ is convariantly constant in $(\M_{(D-2)},h_{ij})$.

\subsection{Odd dimensions}

For odd $D$, as above one can reduce the line element to the form~(\ref{geo_metric fin}), (\ref{general form of H}). Since in odd dimensions ${\etta=0}$ identically (i.e. ${\mbox{\boldmath$\tilde F$}}=0$, see (\ref{spacefield})), a complete solution of the Maxwell equations is now simply given by a purely electric ``radial'' field
\begin{eqnarray}
  \mbox{\boldmath$F$} = \frac{\q}{r^{D-2}} \d r \wedge \d u \qquad (D \mbox{ odd}) .
\end{eqnarray}
As in even $D$ with $\etta=0$, these are again a generalization of the familiar Reissner--Nordstr\" om--de~Sitter spacetimes \cite{Tangherlini63,GibWil87}: the standard Schwarzschild-type form 
\be
 \d s^2=-2H(r)\,\d t^2+\frac{\d r^2}{2H(r)}+r^2\,h_{ij}(x)\,\d x^i\d x^j  ,\qquad \mbox{\boldmath$F$} = \frac{\q}{r^{D-2}} \d r \wedge \d t,
 \label{geo_metric fin_diag}
\ee
is achieved via the transformation\footnote{This transformation (which also applies in the even dimensional case with $F\neq 0$) explicitly shows that the two WANDs~(\ref{nulldirections}) are related by ``time reflexion'', as observed for arbitrary algebraically special static spacetimes in \cite{PraPraOrt07}. This also implies that the two WANDs must have equivalent optical properties (e.g., geodeticity). Note, in particular, that while $\mbox{\boldmath$k$}$ is a principal null direction of the Maxwell 2-form by construction, it turned out that also $\mbox{\boldmath$l$}$ shares this property.}  ${\d u=\d t-\d r/2H}$. Recall also that these represent the only Robinson--Trautman solutions with an aligned electromagnetic field which obeys either the Maxwell or the Maxwell--Chern--Simons equations.

\subsection{The special case ${D=6}$}

The even dimensional ${D=6}$ case is special in that $\etta^2$ may depend also on the spatial $x$ coordinates, see eq.~(\ref{eta2i}). This fact has two consequences. First, one has to solve the more complicate eq.~(\ref{special6}). In addition, when ${\q=0}$, one can not conclude now that $P(x,u)$ takes the factorized form ${ P(x,u) = P(x) \, U(u)}$. Let us discuss the two possible cases separately.

\subsubsection{Factorized $P(x,u)$ (generic transverse space)}

Because of eqs.~(\ref{id_phi}), this corresponds to the generic situation with $\q\neq 0$. In such a case we can arrive again at~(\ref{factorizedP}) so that ${h_{ij}=h_{ij}(x)}$, and eqs.~(\ref{id_phi}) and (\ref{logarithmic derivative of eta}) lead to 
\be 
 \q = \mbox{const}, \qquad \etta^2=\etta^2(x) . 
\ee 
In addition, eq.~(\ref{special6}) simplifies to ${\mu_{,u}(u)=-\frac{1}{16}\Delta(\etta^2)(x)}$ which requires \emph{both terms to be constant}. By integration we obtain ${\mu(u)=\mu_0+c_0u\>}$ and ${\>\Delta(\etta^2)(x)=-16c_0}$, where ${\mu_0, c_0}$ are constants. 

If we restrict to the case when $(\M_{(4)},h_{ij})$ is compact, as for black hole solutions, by standard results (cf., e.g., \cite{Bochner48}, and \cite{KobNom2} on p.~338) the only regular solution is
\begin{equation}
  \mu = \mbox{const}, \qquad \etta^2= \mbox{const} , 
\end{equation}
as in the $D>6$ even dimensional case. Therefore the results of subsection~\ref{subsec_evengen} apply, and $(\M_{(4)},h_{ij},J^i_{\, j})$ is again  (almost-)K\"ahler Einstein (e.g. flat, or $S^2\times S^2$, etc.).

\subsubsection{Non-factorized $P(x,u)$ (transverse space of constant curvature)}

From (\ref{id_phi}) we observe that this case is possible only for $\q=0$, so that we can assume $\etta\neq 0$ (otherwise the Maxwell field would be identically zero). When $P(x,u)$ is non-factorized, as in \cite{Ortaggio07} one can argue that the Riemannian metric $h_{ij}(x,u)$ describes a family of conformal 4-dimensional Einstein spaces parametrized by $u$. It is well known that four-dimensional Riemannian Einstein spaces which admit a conformal (non-homothetic) map on Einstein spaces \emph{must be of constant curvature} \cite{Brinkmann24}. Since $h_{ij}=P^{-2}(x,u)\gamma_{ij}(x)$, this means that we can always find suitable $x$ coordinates such that 
\be 
 h_{ij}=P^{-2}\delta_{ij} , \qquad P=a(u)+b_i(u)\,x^i+c(u)\,\delta_{ij}x^ix^j . \label{6flatP}
\ee
Here ${i,j=1,\ldots,4}$, and ${a(u), b_i(u), c(u)}$ are arbitrary functions of $u$ related to the constant curvature $K$ by $K=4ac-\sum_{i=1}^{4} b_i^2$ \cite{PodOrt06}. Recall also that $(\M_{(4)},h_{ij},J^i_{\, j})$ must be almost-Hermitian. The self-dual ``spatial'' Maxwell field (cf.~(\ref{spec_Max}) with $n=2$) is proportional to the almost-complex structure and must satisy the Maxwell equations~(\ref{constrijF1}) (or, now equivalently, (\ref{nc5})). In addition, there is the constraint~(\ref{special6}). 

Using~(\ref{eta2}), the (analogue of the) Ricci identity applied to the 2-form $F_{ij}$, the effective Maxwell equations (\ref{constrijF1}), (\ref{nc5}), and the constant curvature equation ${\cal R}_{ijkl}=K(h_{ik}h_{jl}-h_{il}h_{jk})$, we can write (cf., e.g. \cite{Bochner48}, for detailed calculations) $\Delta(\etta^2)=2(4K\etta^2+h^{mi}h^{nj}h^{pk}F_{ij||k}F_{mn||p})$ or, by (\ref{eta2}) and (\ref{6flatP}),
\be
 \Delta(\etta^2)=2P^4\Big(4K\sum_{i,j=1}^{4}F_{ij}F_{ij}+P^2\sum_{i,j,k=1}^{4}F_{ij||k}F_{ij||k}\Big) .
 \label{laplace_constr}
\ee

The right hand side of eq.~(\ref{laplace_constr}) is non-negative for $K\ge0$. Therefore, if we again restrict to the case of a {\em compact} $(\M_{(4)},h_{ij})$, for $K\ge0$ standard results \cite{Bochner48,KobNom2} imply that $\etta^2$ does not depend on the $x$ coordinates, and that $F_{ij||k}=0$. In particular, the case $K>0$ requires also $\etta^2=0$, i.e. $F_{ij}=0$ and there is no electromagnetic field (we had already $Q=0$). For $K=0$, as in subsection~\ref{subsec_evengen} thanks to $F=F(u)$ we can achieve $P=P(x)$ (and $\etta=\mbox{const}$, $\mu = \mbox{const}$). But now one can rescale and shift the spatial coordinates to fix $P=1$, i.e. $h_{ij}=\delta_{ij}$ is manifestly flat and $F_{ij||k}=0$ becomes $F_{ij,k}=0$ (this is the solution of subsubsection~\ref{subsubsec_example} with $\epsilon=0$, $D=6$).

The exceptional case $\etta^2=\etta^2(x,u)$, $P=P(x,u)$ (non-factorized) can thus possibly arise only when the transverse space $(\M_{(4)},h_{ij})$ is non-compact, or of constant negative curvature $K=-1$ (in which case eq.~(\ref{laplace_constr}) does not prevent it from being compact, in principle, provided now $F_{ij||k}\neq0$). We do not investigate further this very special case here. Let us only observe that $\pa_u$ is no longer a Killing vector field since the metric depends on $u$.

\subsection{Inclusion of pure radiation}

It is not difficult to generalize these results to include a pure radiation field aligned with the null vector~{\boldmath $k$}. In that case, the total energy-momentum tensor to insert into the Einstein equations is given by the sum of the electromagnetic energy-momentum tensor (\ref{Energy momentum}) and the pure radiation contribution $\tilde T_{\alpha\beta}=\Phi^2 k_\alpha k_\beta$. In the coordinate system introduced above this means that only the ${\tilde T_{uu}=\Phi^2 }$ component is non-vanishing. Moreover, since the covariant divergence of the electromagnetic energy-momentum tensor (\ref{Energy momentum}) vanishes, the Bianchi identities imply ${\tilde T^{\alpha\beta}_{\quad ;\beta}=0}$. For the Robinson--Trautman family of spacetimes this leads to (cf.~\cite{PodOrt06})
\be
\Phi^2=r^{2-D}n^2(x,u)\,,
\ee
where $n$ is an arbitrary function of $x$ and $u$. 

This additional term modifies the field equation of subsection~\ref{subsec_Ruu}. Instead of (\ref{logarithmic derivative of mu}), in the generic case we obtain the equation
\be
 (D-1)\,\mu\,(\ln P)_{,u}-\mu_{,u} =\frac{16\pi\, n^2}{D-2} \qquad (D\not=6) \,. \label{constrspec2pure}
\ee
It is thus possible to \emph{prescribe} the ``mass function'' $\mu(u)$, and the relation (\ref{constrspec2pure}) then uniquely determines the corresponding null matter profile $n^2(x,u)$, provided its left hand side is positive. In the exceptional case ${D=6}$ the equation (\ref{special6}) becomes
\be
  5\,\mu\,(\ln P)_{,u}-\mu_{,u}-{\textstyle\frac{1}{16}}\Delta(\etta^2) =4\pi\, n^2 \qquad (D=6) \label{special6pure} .
\ee 
Again, when the left hand side is positive, this may be considered as the definition of the function~$n$. 
We do not study further details of pure radiation spacetimes here. Let us just observe that purely electric solutions (such that ${\etta=0}$) contain generalized charged Vaidya spacetimes, cf. \cite{ChaBhuBan90}.

\section{Conclusions}
\label{sec_conclusions}


We have derived systematically all higher dimensional spacetimes that contain a hypersurface orthogonal, non-shearing and expanding congruence of null geodesics, together with an aligned electromagnetic field. These are solutions of the coupled Maxwell(--Chern--Simons) and Einstein equations (for any value of the cosmological constant). As already noticed in the vacuum case \cite{PodOrt06}, there appear important differences with respect to the standard ${D=4}$ family of Robinson--Trautman solutions \cite{RobTra60,RobTra62,Stephanibook}. In particular, for ${D>4}$ there is no analogue of radiative spacetimes such as the charged C-metric, and aligned null Maxwell field are not permitted. After integrating the full set of equations, one is essentially left only with (a variety of) static black holes (exceptional subcases possibly arise in $D=6$). These are characterized by mass, electric charge and cosmological constant, and by the topology and geometry of the horizon, which must be an Einstein space. In even spacetime dimensions an additional magnetic parameter is permitted provided the horizon is not only Einstein but also (almost-)K\"ahler. Some of the presented solutions were already known (see the references mentioned above), but we have obtained them systematically as elements of the Robinson--Trautman class, which was the purpose of our work.

Our contribution also makes contact with recent studies of the algebraic classification of the Weyl tensor and of geometric optics in higher dimensions. For instance, it has been recently shown \cite{PraPraOrt07} that arbitrary ${D>4}$ {\em static} spacetimes can be only of the algebraic types G, I$_i$, D or O. Using another result of \cite{PraPraOrt07}, we have demonstrated that our specific static solutions are restricted to the type~D, and we have also given the corresponding WANDs with no need to compute the Weyl tensor. In addition, along with various previous results \cite{FroSto03,Pravdaetal04,PodOrt06,OrtPraPra07,PraPraOrt07}, the new features pointed out above for ${D>4}$ indicate that in some cases the shear-free assumption might be too strong for expanding solutions in higher dimensions. In future work it would thus be worth investigating spacetimes with shear and expansion, at least with some alternative simplifying assumptions.

\section*{Acknowledgments}

We are grateful to Alessio Celi for useful discussions. Part of the work of M.O. was carried out at Dipartimento di Fisica, Universit\`a degli Studi di Trento, whereas his stay in Barcelona has been supported by Fondazione Angelo Della Riccia (Firenze). J.P. and M.\v Z. have been supported by the grant GA\v{C}R~202/06/0041, and by the Czech Ministry of Education under the projects MSM0021610860 and LC06014. M.\v Z. has been further supported by the grant GA\v{C}R~202/05/P127.

\section*{Appendix. The special case of $D=4$}
\renewcommand{\theequation}{A\arabic{equation}}
\setcounter{equation}{0}

For comparison, we will present here a summary of the results in the familiar case ${D=4}$ \cite{RobTra62,Stephanibook}.  We first note that the trace (\ref{trace}) of the energy-momentum tensor of the electromagnetic field is now zero, ${T_\mu^{\, \mu}=0}$. Maxwell's equations still imply (\ref{Fij})--(\ref{Fui}), which now read 
\be
  F_{ij} = F_{ij} (x,u), \qquad
  F_{ru} =r^{-2} \q(x,u), \qquad
  F_{ui}= r^{-1} \q_{,i} - \xi_i (x,u) , \label{AFui}
\ee
where  ${i,j=1,2}$, so that $F_{12}$ is the only independent $F_{ij}$ component. 
The invariants of the Maxwell field are thus
\begin{equation}
  F_{\mu \nu} F^{\mu\nu} = r^{-4} (\etta^2 - 2 \q^2), \qquad {F_{\mu \nu}} {\,{^*}F^{\mu\nu}} = 4r^{-4}P^2F_{12} \q, \label{AF_invariant}
\end{equation}
so that there can be null Maxwell fields when ${\q=0=\etta}$ (i.e., ${F_{ru}=0=F_{ij}}$ while ${F_{ui}=-\xi_i}$).
The source-free equation (\ref{F_ijk}) is now an identity.  We further have the relation (\ref{F_ijuN}).
Finally, the remaining Maxwell equations (\ref{eigen_j}), (\ref{eigen_u}), when expanded in powers of $r$ using previous results, yield the following set of relations:
\begin{eqnarray}
F_{jk}f^k   &=& \q\,h_{jk}f^k, \label{Ad2}\\
 \sqrt{h}\,h^{ij}\,\q_{,j} &=& (\sqrt{h}\,h^{ik}h^{jl}F_{kl})_{,j}\>, \label{Ad3}\\
 (\sqrt{h}\,\q)_{,u}-(\sqrt{h}\,\q\,e^i)_{,i} &=&  \Big(\sqrt{h}\,h^{ij}( \xi_j-F_{jk}e^k )\Big)_{,i}\>. \label{Ad5}
\end{eqnarray}
Note that the remaining condition $(\sqrt{h}\,h^{ij}\,\q_{,j})_{,i} = 0$ is satisfied identically as a consequence of (\ref{Ad3}) and the antisymmetry of $F_{kl}$.

Applying now the field equation for $R_{ij}$, we observe that the powers of $r$ in (\ref{grr_old}) coincide in the terms corresponding to $c_4$, $c_7$, and $c_8$. The expansion of $g^{rr}$ then only contains the first of these terms, yet (\ref{constr_c5}) remains unchanged so we obtain ${f^i=0}$, and we can again set ${e^i=0}$. The expression for $c_4$ is thus modified to ${c_4=\q^2+\etta^2/2}$.
We further find that (\ref{constr_c1}) and (\ref{constr_c7}) remain unchanged. However, for ${D=4}$, they are both identically satisfied so they do not provide additional constraints on $h_{ij}$ and on the electromagnetic field. Thus the expansion of $g^{rr}$ is the same as in (\ref{grr}) but the last two terms are combined (cf. also footnote~\ref{footn_duality}).

Let us also emphasize that in the ${D=4}$ case the spatial metric $h_{ij}$ is 2-dimensional, so that it can always be written in the conformally flat form ${h_{ij}=P^{-2}(x,u)\,\delta_{ij}}$, with ${\sqrt{h}=P^{-2}}$. In fact, for ${D=4}$ we can achieve this by a transformation ${x^i=x^i(\tilde x)}$ involving only the spatial coordinates~$x$, since the $u$-dependence is factorized out as in eq.~(\ref{hijs}). Consequently,  ${{\cal R}=2\Delta\ln P=2P^2[(\ln P)_{,11}+(\ln P)_{,22}]}$.

We can thus summarize that the Robinson--Trautman metric in ${D=4}$ can be cast in the form
\be
 \d s^2=r^2P^{-2}(x,u)\left((\d x^1)^2+(\d x^2)^2\right)-2\,\d u\d r-2H\d u^2 ,
 \label{geo_metric D=4}
\ee
and the aligned electromagnetic field is given by
\be
 \mbox{\boldmath$F$}=\frac{Q}{r^2}\,\d r\wedge\d u+\left(\frac{\q_{,1}}{r} - \xi_1\right)\d u\wedge\d x^1+\left(\frac{\q_{,2}}{r} - \xi_2\right)\d u\wedge\d x^2+F_{12}\,\d x^1\wedge \d x^2 .
\ee 
The various functions and parameters above are constrained by the conditions
\be
2H  =  \frac{{\cal R}}{2}-2\,r(\ln P)_{,u}-\frac{\Lambda}{3}\,r^2-\frac{\mu}{r} +  \frac{\q^2+\frac{1}{2}\etta^2}{r^2} \> ,
  \label{grrD=4}
\ee
where $\mu={\mu(x,u)}$, ${\frac{1}{\sqrt 2}F=P^2F_{12}}$, and (from (\ref{Ad3}), (\ref{Ad5}) and (\ref{F_ijuN}))
\begin{eqnarray}
 &&\textstyle{ \q_{,1}=  (\frac{1}{\sqrt 2}F)_{,2}\>, \qquad
 \q_{,2}  = -(\frac{1}{\sqrt 2}F)_{,1}\>,} \label{Ad4s}\\
 &&\textstyle{ (\q\, P^{-2})_{,u} =\xi_{1,1}+\xi_{2,2}\>, \qquad
  (\frac{1}{\sqrt 2}F\, P^{-2})_{,u}=  \xi_{1,2} - \xi_{2,1}\>.} \label{AF_ijuN}
\end{eqnarray}
Unlike in higher dimensions (cf.~eqs.~(\ref{c3}) and~(\ref{c4})), in ${D=4}$ we have $Q(x,u)$ depending on the spatial coordinates $x$, and $\xi_i(x,u) \not= 0$.

The field equation for $R_{ur}$ in now satisfied. Also the equation (\ref{Ri}) for $R_{ui}$  is satisfied identically, so that one can have a much more general function ${\cal R}(x,u)$. Using (\ref{Ad4s}),  $R_{ui}$ yields only two remaining equations
\begin{eqnarray}\label{D4 mu}
 {\textstyle \mu_{,1} =   4(Q\,\xi_1 - \frac{1}{\sqrt 2}F\, \xi_2) \> , \qquad
  \mu_{,2} = 4(Q\,\xi_2 + \frac{1}{\sqrt 2}F\, \xi_1) }\>.
\end{eqnarray}

Finally, the field equation $R_{uu}$ gives
\begin{eqnarray}
\left(Q^2 + {\textstyle\frac{1}{2}}F^2\right)_{,11}+ \left(Q^2 + {\textstyle\frac{1}{2}}F^2\right)_{,22}   & = & 4(Q_{,1}^2+Q_{,2}^2) \>,\label{D4 Ruu r3}\\
 P^2 (\mu_{,11}+\mu_{,22})+8 (\ln P)_{,u} \left(Q^2 + {\textstyle\frac{1}{2}}F^2\right) - 2 \left(Q^2 + {\textstyle\frac{1}{2}}F^2\right)_{\!,u} & = & 8P^2(Q_{,1}\xi_1+Q_{,2}\xi_1)  \>,\label{D4 Ruu r4}
\end{eqnarray}
and
\be
\Delta{\mathcal R}+12\mu(\ln P)_{,u}-4\mu_{,u} = 4P^2(\xi_1^2+\xi_2^2)\>,
 \label{ARTD4}
\ee
where $\Delta$ is the covariant Laplace operator on a 2-space with metric $h_{ij}$, i.e. ${\Delta{\mathcal R}=P^2({\mathcal R}_{,11}+{\mathcal R}_{,22})}$.

We have thus recovered the well-known results summarized in Theorems~28.3 and 28.7 of~\cite{Stephanibook} with the identification ${\zeta=\frac{1}{\sqrt2}(x^1+\hbox{i}\, x^2)}$, ${h(\zeta,\bar\zeta, u)=\frac{1}{\sqrt 2}(\xi_1+\hbox{i}\,\xi_2)}$, and the complex 
function $Q(\zeta,u)$ related to $Q(x,u)$ and $\frac{1}{\sqrt 2}F(x,u)$ as its real and imaginary parts, respectively. Indeed, (\ref{Ad4s}) are the Cauchy--Riemann conditions so that $Q(\zeta,u)$ must be analytic in $\zeta$. Consequently, $Q_{,11}+Q_{,22}=0=F_{,11}+F_{,22}$, and (\ref{D4 Ruu r3}) is an identity. Equations (\ref{AF_ijuN}) and (\ref{D4 mu}) correspond to equations (28.37e) in reference~\cite{Stephanibook}, (\ref{D4 Ruu r4}) leads to (28.37d), and (\ref{ARTD4}) is exactly the equation (28.37c) in~\cite{Stephanibook}. 

Recall that for $D=4$ electrovacuum Robinson-Trautman solutions can be of the Petrov types II, D or III \cite{Stephanibook}.

%

\providecommand{\href}[2]{#2}\begingroup\raggedright\endgroup

\end{document}